\documentclass[aps,prb,preprint]{revtex4}
\usepackage{graphicx,bm}
\usepackage{amsmath}
\usepackage{color}

\begin{document}
\title{Continuous two-photon source using a single quantum dot in a photonic crystal cavity}
\date{\today}
\author{Harmanpreet Singh, Manoj Das and P. K. Pathak}
\address{School of Basic Sciences,
Indian Institute of Technology Mandi, Kamand, H.P. 175005, India}
\begin{abstract}
We propose methods for realization of continuous two photon source using coherently pumped quantum dot embedded inside a photonic crystal cavity. We analyze steady state population in quantum dot energy levels and field inside the cavity mode. We find conditions for population inversion in coherently pumped and incoherently pumped quantum dot. We show that squeezing in the output for two two photon laser is not visible using coherent as well as incoherent pump. We discuss effect of phonon coupling using recently developed polaron transformed master equation at low temperatures. We also propose scheme for generating squeezed state of field using four wave mixing.
\end{abstract}
\maketitle
\section{Introduction}
In the quest of scalable on-chip quantum technology, semiconductor quantum dots (QDs) have emerged as a potential candidate\cite{chip,chip2,chip3}. With the advanced lithography techniques, it is possible now to grow a quantum dot at desired location inside a photonic crystal microcavity\cite{hennessy, thon}. As a result, new solid state on-chip  cavity quantum electrodynamics (cavity-QED) systems have been developed. The strong coupling regime, where dipole coupling strength between single photon and single QD becomes larger than the damping rates in the system, have been realized\cite{strongcoupling}. Various other phenomena, well-cherished in microwave and optical cavity-QED systems using trapped or Rydberg atoms,  such as appearance of  higher rungs of Jaynes-Cummings ladder\cite{ladder}, photon blockade\cite{photonblockade}, Mollow triplets\cite{mollowtriplet} have also been observed. Further a remarkable success have been achieved in generating sources of nonclassical light such as sources of entangled photons\cite{entangledphoton} and single photon sources having high efficiency and indistinguishability\cite{singlephotonsource}. However, being solid state devices, interactions with longitudinal acoustic phonons are unique in these semiconductor cavity-QED systems. Interactions between phonon and exciton lead to dephasing \cite{phonondephasing} in coupled dynamics of  exciton-photon interaction as well as off-resonant cavity mode feeding\cite{cavitymodefeeding}. Various new phenomena such as high fidelity generation of exciton and biexciton states\cite{biexciton1,biexciton2}, phonon assisted population inversion in two level systems\cite{inversion}, appearance of new features in Mollow triplets\cite{mollow} have been observed due to phonon interactions. Interactions between phonons and excitons play particularly significant role in off resonant exciton-photon interactions\cite{majumdar}.

Generating coherent light by placing a large number of emitters inside a cavity has been fascinating subject and different kinds of lasers have been developed\cite{semiconductorlaser}. In conventional lasers a large number of emitters are used as gain medium to overcome photon losses from the cavity. However, microlasers where a single emitter acts as the gain medium inside high quality cavity, have  also been realized\cite{microlaser}. These systems are particularly useful for the applications in quantum information processing. These lasers operate in the limit where semiclassical laser theory becomes inapplicable and complete quantum theories have been developed. Application of QDs in conventional laser as gain medium has limited success due to variation in their sizes and exciton resonance frequencies\cite{dotsize}. However, the single emitter micro-lasers using single QDs embedded inside high quality photonic crystal cavities\cite{singledotlaser,singledotlaser2} and QD coupled with coplanar microwave cavity\cite{cplanner1, cplanner2} have been realized recently. Further, it has been observed that the phonon interactions significantly alter the dynamics in both type of microlasers, i.e., a single QD embedded in photonic crystal cavity\cite{singledotlaser2} and a single QD coupled with coplanar cavity\cite{cplanner2}, and a largely enhanced output power has been achieved due to phonon assisted off-resonant transitions.

Similar to the single photon lasers where lasing occurs due to stimulated emission of single photons, it was also predicted that a similar coherent generation of light is also possible through stimulated two-photon processes\cite{twophotonlasing}. Further, the output of a two-photon laser would be squeezed coherent state that exhibits quantum properties\cite{twophotonsqueezing}. Although a less success in realization of two-photon laser had been achieved before realization of two-photon mazer\cite{twophotonmaser}. The two-photon laser has been demonstrated  by D. J. Gauthier et al \cite{gauthier} using strongly driven two level atoms as gain medium and probing by a weak field having frequency resonant to one of the side band. The absence of squeezing in the output has been explained due to enhancement of noise in spontaneous generation of photons. However, it was also predicted that two-photon correlated emission laser can show squeezing for certain parameters when the noises in one photon emission cancel each other\cite{scully}. We notice that two photon lasing in a single QD using incoherent pumping has been proposed recently\cite{tejedor}. Here we propose a scheme for realization of two-photon laser using single QD embedded in a photonic crystal cavity. The two-photon emission through cavity mode is dominating in the case when single photon transitions are far off-resonant and two-photon resonance conditions are satisfied\cite{ota}. Since the photon pair is generated through biexciton decay into the far off-resonant cavity mode, the phonon interaction can play a significant role. We use recently developed master equation techniques to include such interactions\cite{masterequation}. In this work we investigate effects of coherent nature of pump in order to achieve correlated two-photon emission in a single QD two-photon laser to predict nonclassical features in laser output.

Our paper is organized as follows. In Sect.~\ref{Sec:incoherent}, we present model for two-photon laser using incoherent pump. We discuss effects of phonon interaction at low temperature in terms of
steady state population dynamics and Wigner function of the cavity field.
In Sect.~\ref{Sec:coherent}, we discuss two-photon lasing using coherent pump and continuous generation of nonclassical field using four-wave mixing is discussed in Sect.~\ref{Sec:fourwave}.
Finally, conclusions are presented in Sect.~\ref{Sec:Conclusions}.
\section{Two-photon lasing using incoherent Pump}
\label{Sec:incoherent}
\begin{figure}[t!]
\centering
\includegraphics[height=2.2in]{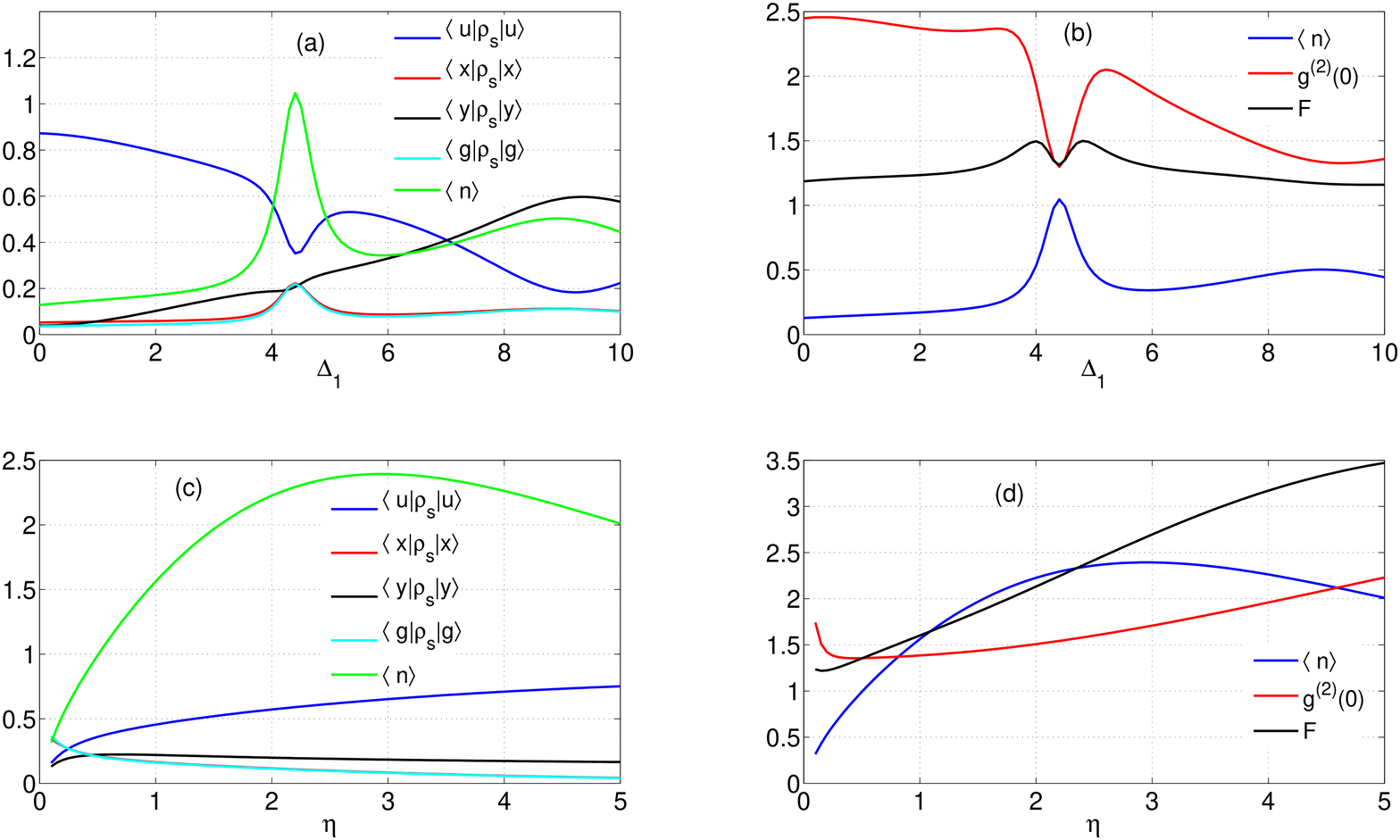}
\vspace{-0.1cm}
\caption{(Color online) Steady state populations in quantum dot energy states and cavity field parameters for temperature $T=5K$, cavity leakage $\kappa = 0.2$, cavity field couplings $g_1 = g_2$, spontaneous decay rates $\gamma_1=\gamma_2 = 0.01g_1$, pure dephasing rate $\gamma_d = 0.01$, biexciton binding energy $\Delta_{xx}=10.0g_1$, anisotropic energy gap $\delta_x$ =1.0. The other parameters are
for (a) \& (b) $\eta_1=\eta_2=0.5g_1$, and for (c) \& (d) $\Delta_1=4.4g_1$.}
\label{fig1}
\end{figure}

\begin{figure}[t!]
\centering
\includegraphics[height=2.2in]{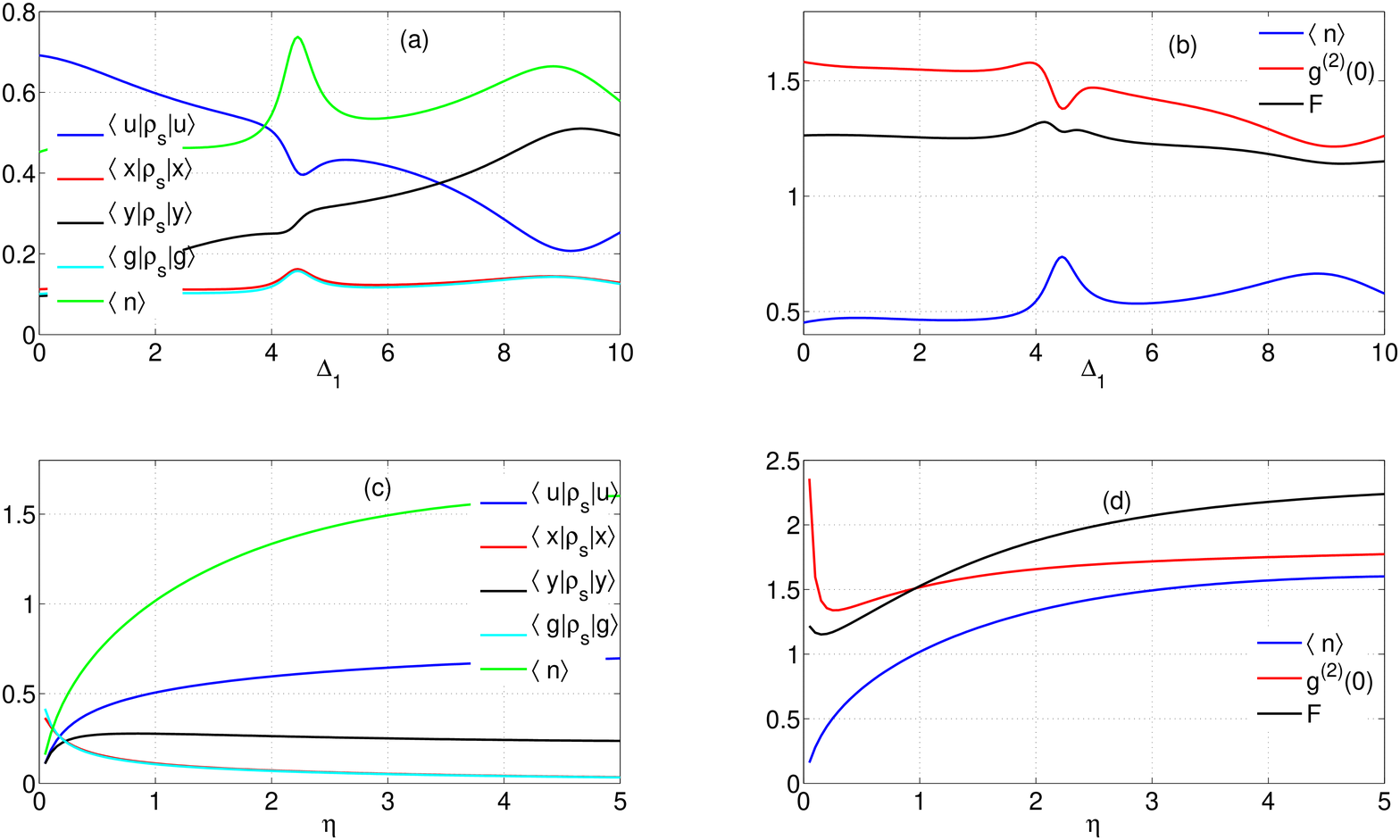}
\vspace{-0.1cm}
\caption{(Color online)
Steady state populations in quantum dot energy states and cavity field parameters for temperature $T=20K$ using same parameters as in Fig.1.}
\label{fig2}
\end{figure}
We consider a single quantum dot embedded in a single mode photonic crystal microcavity. The quantum dot consists of four energy level, ground state $|g\rangle$, two exciton states $|x\rangle$ \& $|y\rangle$ and biexciton state $|u\rangle$. The transitions $|g\rangle\leftrightarrow|x\rangle$ and $|x\rangle\leftrightarrow|u\rangle$ are driven by x-polarized pump field and the transitions $|u\rangle\leftrightarrow|y\rangle$ and $|y\rangle\leftrightarrow|g\rangle$ are coupled with a y-polarized cavity mode. In the QD, exciton state $|x\rangle$ and biexciton  state $|u\rangle$ are created through incoherent x-polarized pump field. The directions of polarization of the cavity mode and the pumping fields are chosen perpendicular to each other. We notice that a similar scheme has been proposed by  Elena del Valle et al\cite{tejedor} without considering effect of acoustic phonons. We investigate the effect of exciton-phonon coupling on two-photon lasing in terms of lasing conditions and the properties of laser output. These effects are particularly significant because the two-photon resonant transitions used for pumping as well as for two-photon generation in cavity mode are far off-resonant. The Hamiltonian of the system in rotating frame with cavity mode frequency is given by
\begin{eqnarray}
\label{incoh-H}
H=\hbar(\delta_1+\delta_x)\sigma_{xx}+\hbar(2\delta_1+\delta_x-\Delta_{xx})\sigma_{uu}+\hbar\delta_1\sigma_{yy}\nonumber\\
+\hbar(g_1\sigma_{yg}a+g_2\sigma_{uy}a+H.c.)+H_{ph}.
\end{eqnarray}
Here $\delta_1=\omega_y-\omega_c$ is  detuning between exciton resonance frequency $\omega_y$ and cavity mode of frequency $\omega_c$, $g_1$ \& $g_2$ are dipole coupling constants for the transitions $|y\rangle\leftrightarrow|g\rangle$ \& $|u\rangle\leftrightarrow|y\rangle$ with the cavity mode, $\sigma_{ij}=|i\rangle\langle j|$ are QD operators and $a$ is annihilation operator for photons in cavity mode. The longitudinal acoustic phonon bath and exciton-phonon interactions are included in $H_{ph}=\hbar\sum_k\omega_kb_k^{\dag}b_k+\sum_{i=x,y,u}\lambda_{ik}\sigma_{ii}(b_k+b_k^{\dag})$; where $\lambda_{ik}$ are exciton phonon coupling constants and $b_k$ \& $b_k^{\dag}$ are annihilation and creation operators for kth phonon mode of frequency $\omega_k$. In order to keep exciton-phonon coupling up to all order we use polaron transformed Hamiltonian. The transformed Hamiltonian
$H^{\prime}=e^PHe^{-P}$ with
$P=\sum_{i=x,y,u}\frac{\lambda_{ik}}{\omega_k}\sigma_{ii}(b_k-b_k^{\dag})$; can be written as the sum of terms corresponding to cavity-QD system, phonon bath and system-bath interactions as $H^{\prime}=H_s+H_b+H_{sb}$, where
\begin{eqnarray}
 H_s=\hbar(\Delta_1+\delta_x)\sigma_{xx}+\hbar(2\Delta_1+\delta_x-\Delta_{xx})\sigma_{uu}+\hbar\Delta_1\sigma_{yy}\nonumber\\
 +\langle B\rangle X_g,\\
 H_b=\hbar\sum_k\omega_k b_k^{\dag}b_k,\\
 H_{sb}=\xi_gX_g+\xi_uX_u.
\end{eqnarray}
The polaron shifts $\sum_k\lambda_{ik}^2/\omega_k$, are included in the effective detunings $\Delta_1$ and $\Delta_2$.
The system operators are given by $X_g=\hbar(g_1\sigma_{yg}a+g_2\sigma_{uy}a+H.c.)$, $X_u=i\hbar(g_1\sigma_{yg}a+g_2\sigma_{uy}a-H.c.)$ and every other symbol has same meaning as discussed in earlier chapters. The final form of master equation in terms of reduced density matrix for cavity-QDs coupled system $\rho_s$ is written as\cite{meq}
\begin{eqnarray}
\dot{\rho_s}=-\frac{i}{\hbar}[H_s,\rho_s]-{\cal L}_{ph}\rho_s-\frac{\kappa}{2}{\cal L}[a]\rho_s\nonumber\\
-\sum_{i=x,y}\left(\frac{\gamma_i}{2}{\cal L}[\sigma_{gi}]
+\frac{\gamma_i^{\prime}}{2}{\cal L}[\sigma_{iu}]\right)\rho_s-\sum_{i=x,y,u}\frac{\Gamma_i}{2}{\cal L}[\sigma_{ii}]\rho_s\nonumber\\
-\left(\frac{\eta_1}{2}{\cal L}[\sigma_{xg}]
+\frac{\eta_2}{2}{\cal L}[\sigma_{ux}]\right)\rho_s,
 \label{meq}
\end{eqnarray}
where $\kappa$ is photon leakage rate from the cavity mode and $\gamma_i$, $\gamma^{\prime}_i$ are spontaneous decay rates. The dephasing rates for exciton states are given by $\Gamma_i$ and  $\eta_1$, $\eta_2$ are corresponding to exciton and biexciton pumping rates. The phonon induced processes are given by
\begin{eqnarray}
 {\cal L}_{ph}\rho_s=\frac{1}{\hbar^2}\int_0^{\infty}d\tau\sum_{j=g,u}G_j(\tau)[X_j(t),X_j(t,\tau)\rho_s(t)]+H.c.
\end{eqnarray}
where $X_j(t,\tau)=e^{-iH_s\tau/\hbar}X_j(t)e^{iH_s\tau/\hbar}$, and
$G_g(\tau)=\langle B\rangle^2\{\cosh[\phi(\tau)]-1\}$ and $G_u(\tau)=\langle B\rangle^2\sinh[\phi(\tau)]$. The phonon bath is treated as a continuum
with spectral function $J(\omega)=\alpha_p\omega^3\exp[-\omega^2/2\omega_b^2]$, where the parameters $\alpha_p$ and $\omega_b$ are
the electron-phonon coupling and cutoff frequency respectively. In our calculations we use $\alpha_p=1.42\times10^{-3}g_1^2$ and $\omega_b=10g_1$, which gives $\langle B\rangle=1.0$, $0.90$, $0.84$, and $0.73$ for $T=0K$, $T=5K$, $10K$, and $20K$, respectively. The
system-phonon interactions are included in phonon correlation function $\phi(\tau)$ given by
\begin{eqnarray}
\phi(\tau)=\int_0^{\infty}d\omega\frac{J(\omega)}{\omega^2}
\left[\coth\left(\frac{\hbar\omega}{2K_bT}\right)\cos(\omega\tau)-i\sin(\omega\tau)\right],
\end{eqnarray}
where $K_b$ and $T$ are Boltzmann constant and the temperature of phonon bath respectively.
We solve master equation (\ref{meq}) numerically using quantum optics toolbox\cite{toolbox}. In order to analyze the two-photon lasing we plot steady state populations and cavity field statistics results in Fig.\ref{fig1} and Fig.\ref{fig2} using typical values of parameters which are compatible with experiments and phonon bath temperatures $T=5K$ and $T=20K$, respectively. We choose biexciton binding energy corresponding to $\Delta_{xx}=10g_1$ and by changing cavity frequency the detuning $\Delta_1$ is changed at constant temperature\cite{oxidation}. The cavity assisted two-photon resonance occurs for $\Delta_1\approx(\Delta_{xx}-\delta_x)/2$. For anisotropic energy gap $\delta_x=g_1$ the two photon resonance occurs around $4.5g_1$, which is slightly modified by cavity induced Stark shifts. The single photon resonances can occur when either cavity is resonant to exciton transition, i.e. $\Delta_1=0$ or the cavity is resonant to the biexciton to exciton transition, i.e. $\Delta_1=\Delta_{xx}-\delta_x$. Since we have considered pumping to biexciton state $|u\rangle$, only through transitions $|g\rangle\rightarrow|x\rangle$ \& $|x\rangle\rightarrow|u\rangle$, the resonant photon emission corresponding to single photon resonance at $\Delta_1=0$ does not occur, as exciton state $|y\rangle$ which is coupled through cavity mode remains less populated because biexciton to exciton transition $|u\rangle\rightarrow|y\rangle$ becomes far off-resonant. When detuning $\Delta_1$ is increased from zero, the steady-state population in biexciton state, $\langle u|\rho_s|u\rangle$ calculated after tracing over cavity states, is much larger than populations in other states of QD. Therefore one can easily achieve population inversion in single QD, when incoherent pumping is larger than other losses in the system. For $\Delta_1=4.4g_1$, i.e. when two-photon resonance occurs, a large population from biexciton state is transferred to ground state $|g\rangle$, thus indicating generation of photons in pairs inside cavity mode from $|u\rangle\rightarrow|g\rangle$ via $|y\rangle$. The population in exciton state $|y\rangle$ keep increasing monotonically for larger positive value of $\Delta_1$ and becomes larger than population in biexciton state around single photon resonance at $\Delta_1\approx\Delta_{xx}-\delta_x$. This is due to the fact that on increasing $\Delta_1$, the detuning between cavity mode and biexciton to exciton transition decreases. The mean number of photons in cavity mode also exhibits a sharp peak around two-photon resonance and a broader peak corresponding to phonon assisted single photon transition at $\Delta_1=\Delta_{xx}-\delta_x$. In Fig.1 (b) \& Fig.2(b), mean photon number $\langle n\rangle=\langle a^{\dag}a\rangle$ and the second order photon correlation for zero time delay, $g^2(0)=\langle a^{\dag 2}a^2\rangle/\langle a^{\dag}a\rangle^2$, which indicates two-photon coincidence detection, has been plotted. The Fano factor $F=(\langle n^2\rangle-\langle n\rangle^2)/\langle n\rangle$ defined as the ratio of variance for the field inside the cavity mode to that for the coherent state having same average number of photons, has also been plotted. In Fig.1(b), when detuning $\Delta_1$ increases from zero to the value $(\Delta_{xx}-\delta_x)/2$ corresponding to two-photon resonance, the average number of photon inside cavity mode $\langle n\rangle$ become maximum at the same time the photon correlation function $g^2$ becomes minimum. It has been proved that the value of $g^2$ for two photon coherent states could be larger than $1$\cite{twophotonsqueezing}. The Fano factor $F$ also acquires minimum value, indicating suppression of noise and large enhancement in cavity field simultaneously at two-photon resonance. Therefore two-photon lasing in single QD using incoherent pump is possible. In subplots (c) \& (d) of Figs. 1 \& Fig.2, steady state populations and cavity field statistics has been shown for different values of pump strength, using $\eta_1=\eta_2$ and $\Delta_1$ corresponding to two-photon resonance.  When pump intensity is increased more population in biexciton state start growing and beyond certain threshold value population inversion in achieved. The population inversion grows on further increasing pump power and  saturates after attaining maximum value. The average number of photons inside cavity mode $\langle n\rangle$ also increases first on increasing pump field and becomes maximum at certain pump strength, on increasing pump power further value of $\langle n\rangle$, which is well understood self quenching effect in single emitter lasers.

In Fig.2, the temperature of phonon bath is chosen $T=20K$. Comparing Fig.1 \& Fig.2, we notice that at higher temperature, due to phonon assisted off resonant transitions, the emission into cavity mode is enhanced for all off-resonant values of detuning $\Delta_1$. However, the emission around two-photon resonance reduced slightly and enhanced around single photon resonance at $\Delta_1=(\Delta_{xx}-\delta_x)/2$. Further the maximum values of population inversion and average number of photons inside cavity mode decreases. On increasing temperature the values of $g^2(0)$ and Fano factor $F$ increases under two-photon lasing condition showing enhancement in noise.
\begin{figure}[t!]
\centering
\includegraphics[height=2.2in]{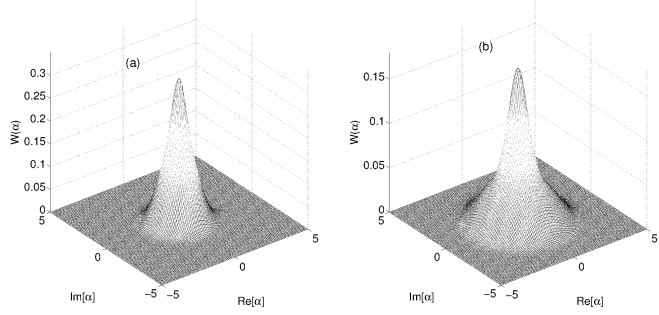}
\vspace{-0.1cm}
\caption{(Color online) Wigner distribution of cavity field for incoherently pumped quantum dot using $\eta_1=\eta_2=0.5g_1$, $\Delta_1=4.4g_1$, for (a) $T=5K$, for (b) $T=20K$ and other parameters same as in Fig.1.} \label{fig3}
\end{figure}
Next, we plot the Wigner distribution for the cavity field in Fig. \ref{fig3}. The Wigner distribution is calculated using following definition\cite{wigner}
\begin{eqnarray}
    W(\alpha)=\frac{2}{\pi^{2}}e^{2|\alpha|^{2}}\int d^{2}\beta\langle-\beta|\rho_c|\beta\rangle e^{-2(\beta\alpha^{*}-\beta^{*}\alpha)}
    \label{wigner1}
\end{eqnarray}
where, $\rho_c=\sum_{n,m}\rho_{nm}|n\rangle\langle m|$, is the density matrix for the cavity field at steady state calculated after tracing over the QD states and $|\beta\rangle$ is a coherent state. Using the density matrix $\rho_c$ in Eq. (\ref{wigner1}),
the Wigner distribution takes the following form
\begin{eqnarray}
    W(\alpha)=\frac{2}{\pi^{2}}e^{2|\alpha|^{2}}\sum_{n,m}\rho_{nm}\int  d^{2}\beta\frac{(-\beta^{*})^{n}\beta^{m}}{\sqrt{n!m!}}e^{-|\beta|^2}\nonumber\\
    \times e^{-2(\beta\alpha^{*}-\beta^{*}\alpha)}.
    \label{wigner2}
\end{eqnarray}
Further, after evaluating the integration in Eq. (\ref{wigner2}), we get
\begin{eqnarray}
    W(\alpha)=\frac{2}{\pi}e^{2|\alpha|^{2}}\sum_{n,m}\frac{\rho_{nm}}{\sqrt{n!m!}}\frac{(-1)^{n+m}}{2^{n+m}}\frac{\partial^{n+m}}{\partial\alpha^{n}\partial\alpha^{*m}}e^{-4|\alpha|^2}.
\label{wigner3}
\end{eqnarray}
In Eq. (\ref{wigner3}), the term $\frac{\partial^{n+m}}{\partial\alpha^{n}\partial\alpha^{*m}}e^{-4|\alpha|^2}$ is calculated using Leibniz rule. The Eq. (\ref{wigner3}) is simplified to
\begin{eqnarray}
    W(\alpha)=\frac{2}{\pi}e^{-2|\alpha|^{2}}\sum_{n,m}\rho_{nm}\sum_{i=0}^n\frac{(-1)^{n-i}\sqrt{n!m!}}{n-i!}\nonumber\\\times\frac{(2\alpha^{*})^{i}}{i!}\frac{(2\alpha)^{m-n+i}}{m-n+i!}.
\label{wigner4}
\end{eqnarray}
In Fig.3(a), we show Wigner distribution (\ref{wigner4}) for parameters used in Fig.1(a) and in Fig.3(b) for parameters used in Fig.2(a) with detuning $\Delta_1=4.4g_1$ corresponding to two-photon lasing. It becomes clear on comparing Fig.3(a) \& (b) that on increasing temperature of phonon bath the variance in cavity field increases. Further, squeezing in cavity field as predicted for two-photon coherent states is absent. The absence of squeezing in two-photon laser has been understood due to the fact that the noise in fields of each photons in the pair get added which negate squeezing in cavity output.
\section{Two-photon lasing using coherent Pump}
\label{Sec:coherent}
For pumping QD in biexciton state coherently we consider an external laser applied between transitions $|g\rangle\rightarrow|x\rangle$ \& $|x\rangle\rightarrow|u\rangle$. The polarization of laser field is chosen orthogonal to the polarization of cavity field. The Hamiltonian of the system in rotating frame with frequency of pump laser $\omega_p$ is given by
\begin{eqnarray}
H=\hbar\Delta_p\sigma_{xx}+\hbar(2\Delta_p-\delta_x-\Delta_{xx})\sigma_{uu}+\hbar(\Delta_p-\delta_x)\sigma_{yy}\nonumber\\
+\hbar(\Delta_p-\delta_x-\Delta_1)a^{\dag}a+\hbar(\Omega_1\sigma_{xg}+\Omega_2\sigma_{ux}+H.c.)\nonumber\\
+\hbar(g_1\sigma_{yg}a+g_2\sigma_{uy}a+H.c.)+H_{ph}.
\label{coh_H}
\end{eqnarray}
Here $\Delta_p=\omega_x-\omega_p$ and $\Omega_i$ are the detuning and dipole coupling strengths for the pump laser, everything else has their previously defined meaning. After following the same procedure we arrive at the same polaron transformed master Eq.(\ref{meq}). Where system Hamiltonian and operators $X_g$ and $X_u$ are defined as follows
\begin{eqnarray}
 H_s=\hbar\Delta_p\sigma_{xx}+\hbar(2\Delta_p-\delta_x-\Delta_{xx})\sigma_{uu}+\hbar(\Delta_p-\delta_x)\sigma_{yy}\nonumber\\
+\hbar(\Delta_p-\delta_x-\Delta_1)a^{\dag}a+\langle B\rangle X_g,\\
X_g=\hbar(\Omega_1\sigma_{xg}+\Omega_2\sigma_{ux}+g_1\sigma_{yg}a+g_2\sigma_{uy}a)+H.c.\\
X_u=i\hbar(\Omega_1\sigma_{xg}+\Omega_2\sigma_{ux}+g_1\sigma_{yg}a+g_2\sigma_{uy}a)+H.c..
\end{eqnarray}
The polaron shifts are absorbed in the detunings.
\begin{figure}[t!]
\centering
\includegraphics[height=2.2in]{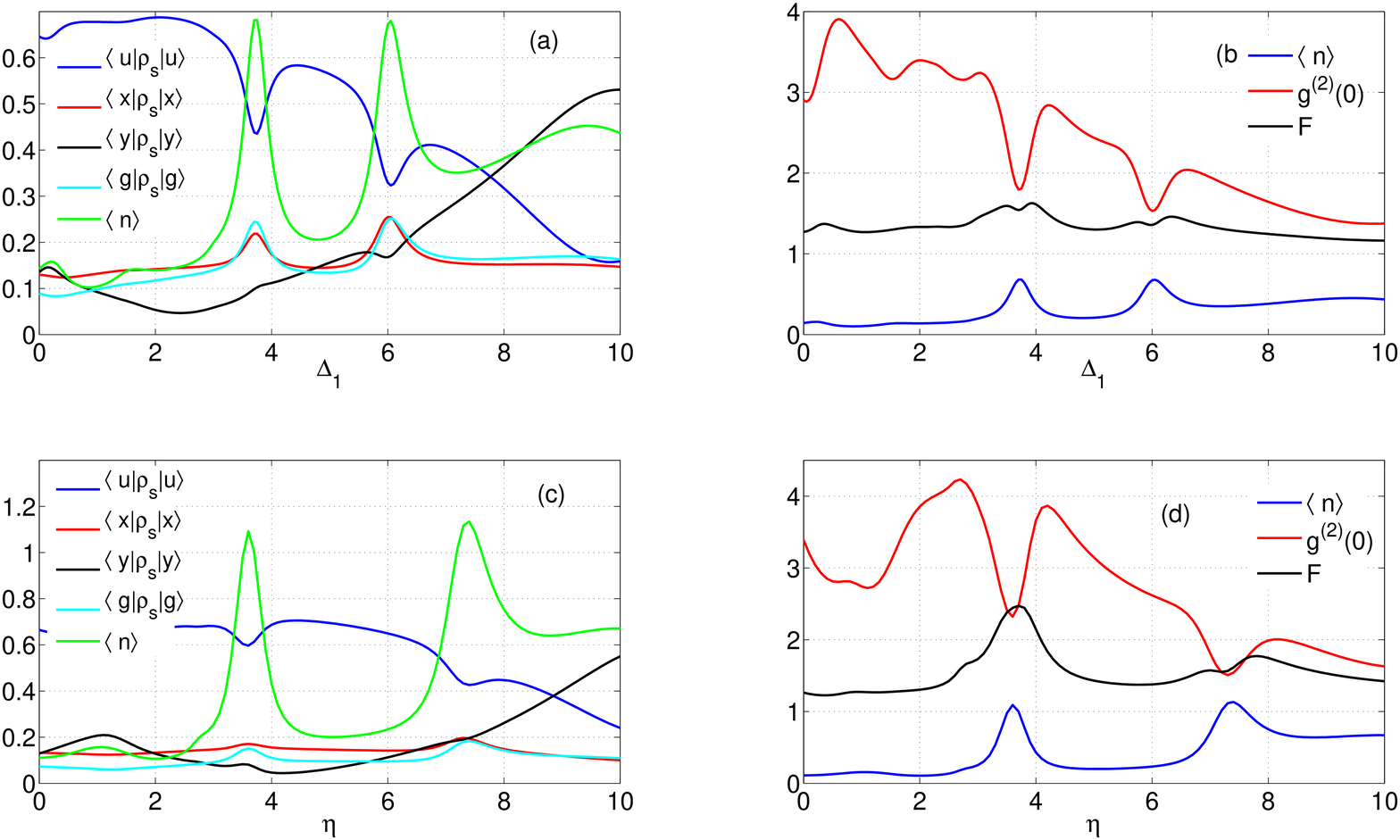}
\vspace{-0.1cm}
\caption{(Color online)Steady state populations in quantum dot energy states and cavity field parameters for coherently pumped quantum dot for $\Delta_p=0$, $\Omega_1=\Omega_2=2.4g_1$ and other parameters are same as in Fig.1.} \label{fig4}
\end{figure}
\begin{figure}[t!]
\centering
\includegraphics[height=2.2in]{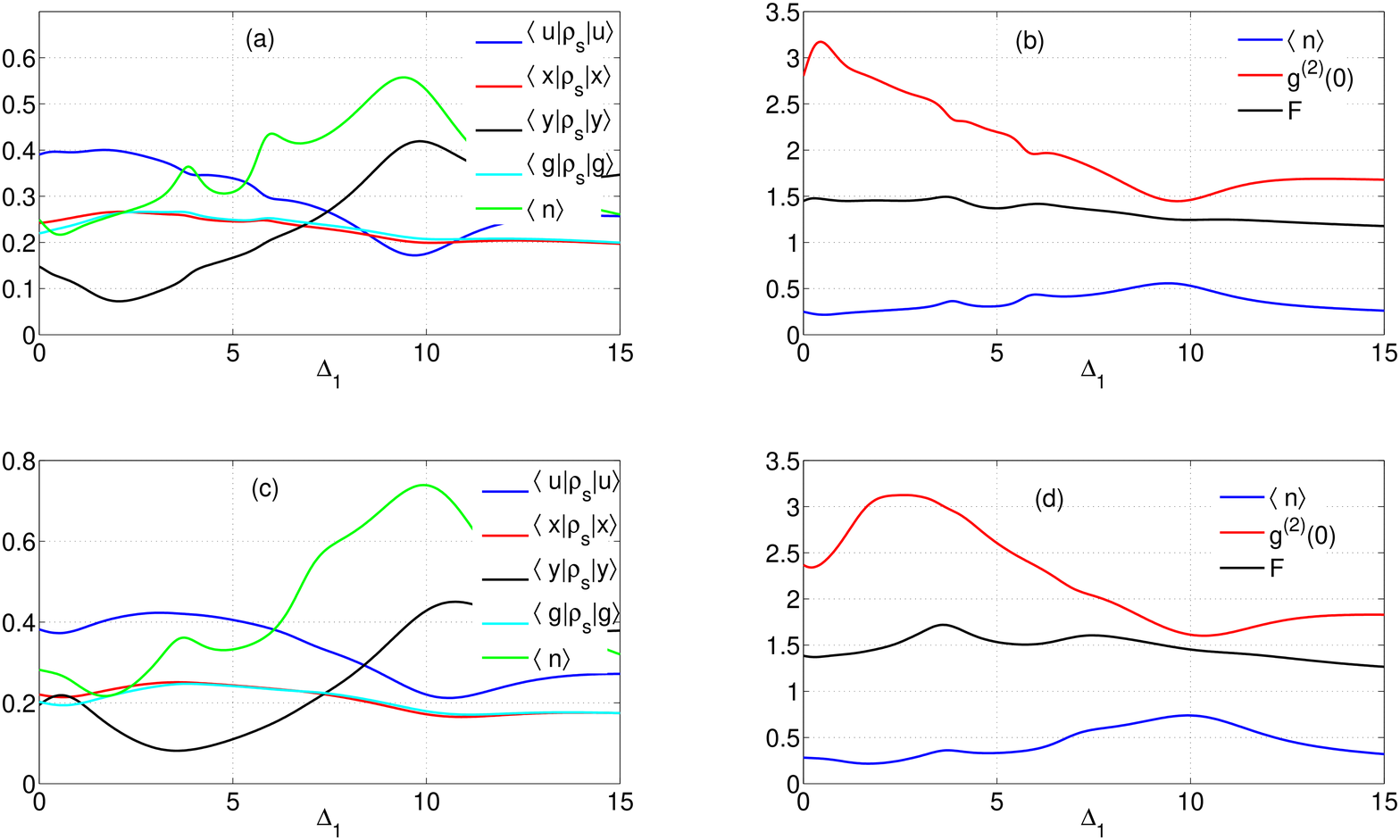}
\vspace{-0.1cm}
\caption{(Color online)Steady state populations in quantum dot energy states and cavity field parameters for coherently pumped quantum dot for $\Delta_p=0$, $\Omega_1=\Omega_2=4.0g_1$ and other parameters are same as in Fig.2.} \label{fig5}
\end{figure}
\begin{figure}[t!]
\centering
\includegraphics[height=2.2in]{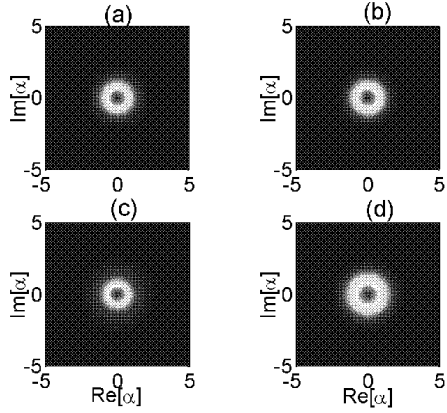}
\vspace{-0.1cm}
\caption{(Color online) Wigner distribution of cavity field for coherently pumped quantum dot using, in (a) \& (c) $\Delta_1=3.7g_1$, in (b) \& (d) $\Delta_1=6.0g_1$. The other parameters in (a) \& (b) are same as in Fig.4 and in (c) \& (d) are same as in Fig.5.} \label{fig6}
\end{figure}

In Fig.4 and Fig.5, we show steady state populations and cavity field statistics at $T=5K$ and $T=20K$ using coherent pump, respectively. For subplots (a) \& (b), we consider pump strength such that $\Omega_1=\Omega_2=2g_1$ and for subplots (c) \& (d), we use $\Omega_1=\Omega_2=4g_1$. For typical values of parameters, we get population inversion for a good range of positive values of $\Delta_1$. However, when $\Delta_1$ becomes comparable to biexciton binding energy single photon transitions from $|u\rangle\rightarrow|y\rangle$ are enhanced and population in $|y\rangle$ start dominating. The pump laser is applied resonantly between $|g\rangle\rightarrow|x\rangle$ transition thus making transition $|x\rangle\rightarrow|u\rangle$ detuned by biexciton binding energy. In the presence of phonon interaction such method of pumping has been found very efficient for deterministic generation of biexciton state\cite{biexpump}. One should notice here phonon interaction is essential for achieving population inversion through this method. Due to resonant application of pump laser between $|g\rangle$ and $|x\rangle$, dressed states $|\pm\rangle=(|x\rangle\pm|g\rangle)/\sqrt{2}$, are formed. Therefore the ground state effectively splits into a doublet separated in frequency by pump laser strength $\Omega_1$. Due to this splitting of ground state into a doublet, two-photon resonant emission in cavity mode occurs corresponding to two values of $\Delta_1$ separated by $\Omega_1$. Since the influence of phonon coupling is more pronounced for larger detunings the two-photon resonance peak at higher values of $\Delta_1$ has larger width. Similar to the case of two-photon lasing using incoherent pumping, we get troughs in the plots of $g^2(0)$ and $F$ for the values of $\Delta_1$ corresponding to two-photon resonant emission. Clearly, sharp rise in average number of photons in cavity mode and the reduction of variance in photon distribution indicate two-photon lasing at two different values of $\Delta_1$. At higher temperature $T=20K$, as shown in Fig.5, the two photon resonance peaks in average number of cavity photons disappear and emission in cavity mode from off-resonant single photon transitions, for all values of $\Delta_1$, is enhanced due to phonon coupling. Although population inversion can be achieved at higher phonon bath temperature but the emission at two photon resonance in cavity is inhibited which restrict two-photon lasing using coherent pump.

In Fig.6, we plot Wigner function calculated using Eq.(\ref{wigner3}) corresponding to parameters used in Fig.4 at two photon resonance. In this case also no squeezing is observed as the procedure of two-photon generation in cavity mode remains same. Further we observe larger broadening in Wigner function corresponding to larger values of $\Delta_1$.
\section{Continuous source of squeezed light through Fourwave Mixing}
\label{Sec:fourwave}
\begin{figure}[t!]
\centering
\includegraphics[height=2.2in]{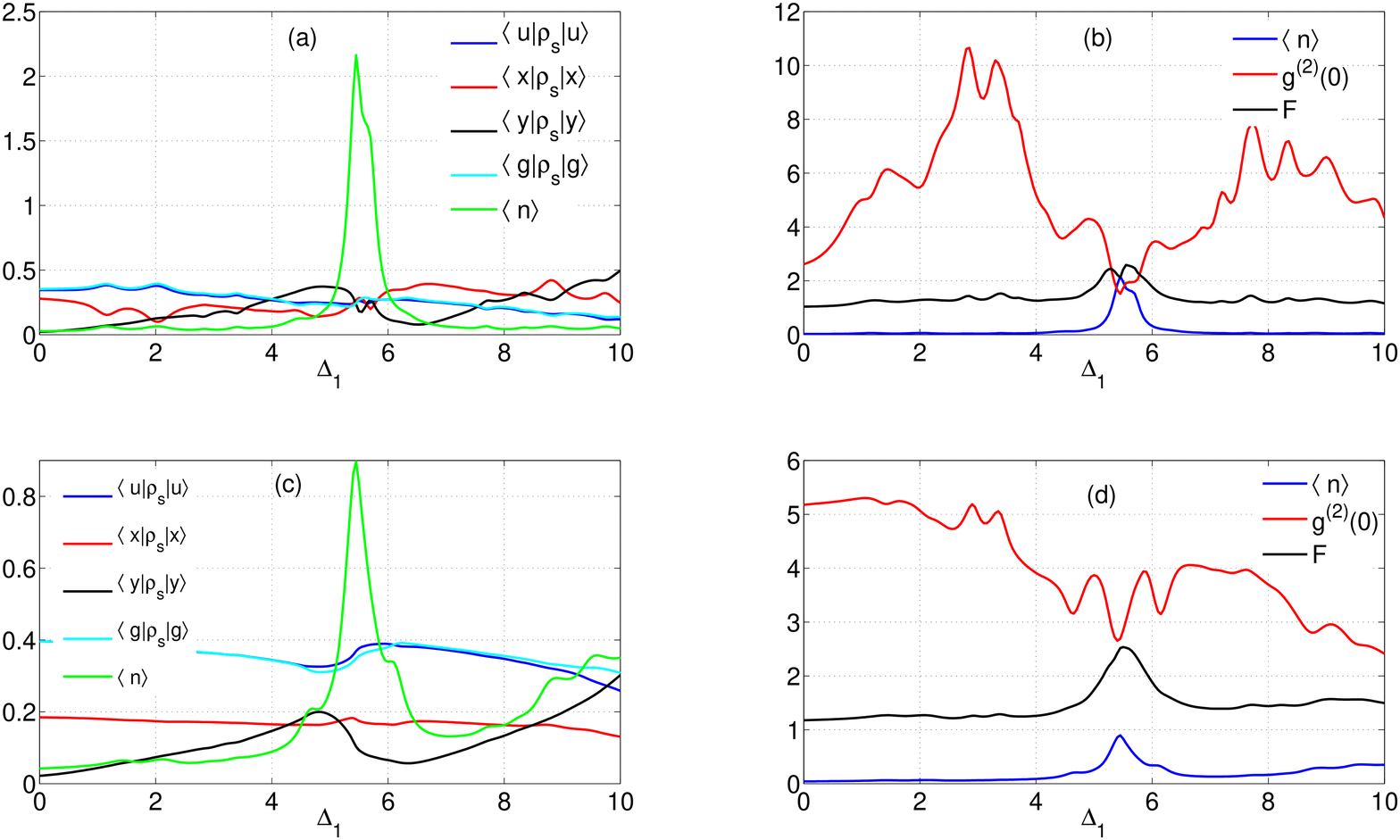}
\vspace{-0.1cm}
\caption{(Color online) Steady state populations in quantum dot energy states and cavity field parameters in fourwave mixing for $\Delta_p=4.5g_1$, $\delta_x=-g_1$, $\Omega_1=\Omega_2=2.5g_1$, for (a) \& (b) temperature $T=0K$, and for (c) \& (d) $T=20K$.} \label{fig7}
\end{figure}

\begin{figure}[t!]
\centering
\includegraphics[height=2.2in]{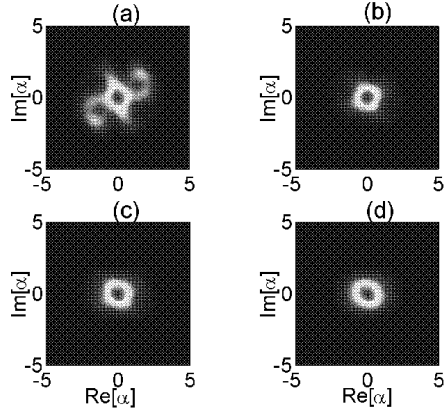}
\vspace{-0.1cm}
\caption{(Color online) Wigner distribution of cavity field in fourwave mixing using same parameters as in Fig.7 and $\Delta_1=5.5g_1$.}
\label{fig9}
\end{figure}
In this case we consider coherent pumping using same setup discussed in previous section. However, the QD is pumped in biexciton state through two-photon resonant transition $|g\rangle\rightarrow|u\rangle$ via exciton state $x\rangle$. We choose detuning of the pump $\Delta_p=(\Delta_{xx}+\delta_x)/2$ which satisfy two-photon resonant condition. The couplings of pump laser $\Omega_1$ and $\Omega_2$ are considered larger than QD-cavity couplings $g_1$ and $g_2$ for stronger pumping.

In Fig.7, we plot steady state population and cavity field parameters. In subplots (a) \& (b), we do not include phonon-exciton interactions and in (c) \& (d) we consider phonon interaction at $T=5K$. In this pumping method we get almost equal populations in ground state $|g\rangle$ and biexciton state $|u\rangle$, which can be dominating. However, no significant population inversion can be achieved. The average number of cavity photons has sharp two-photon resonance peak corresponding to $\Delta_1=(\Delta_{xx}-\delta_x)/2$ and $g^2(0)$ and Fano factor sharp trough. On including phonon-exciton interaction, the value of maximum $\langle n\rangle$ inside the cavity mode decreases and the value of $g^2(0)$ and Fano factor increases, indicating enhancement in variance of photon distribution. In Fig.8, we show Wigner distribution of cavity field corresponding to two-photon resonance and temperature $T=0K$, $T=5K$, $T=10K$ and $T=20K$ in subplots (a), (b), (c), and (d), respectively. At $T=0K$, when exciton-phonon interactions are absent, squeezing as well as interference patterns appear due to coherence in emitted photons. However when temperature is increased and exciton-phonon interactions become significant coherence between emitted photons diminishes these features slowly disappear. However, at low temperature say up to $T=20K$, squeezing in cavity field is visible and one can realize continuous source of squeezed light using single QD.
\section{Conclusions}
\label{Sec:Conclusions}
We have discussed effect of exciton-phonon coupling on two-photon lasing in a single quantum dot embedded inside a photonic crystal cavity. We have analyzed schemes of incoherent as well as coherent pump for achieving two-photon lasing. In order to visualize squeezing in cavity field we have plotted Wigner  function. In the case of two-photon lasing we do not find squeezing in cavity field. However, we discuss method of four-wave mixing for generating continuous source of squeezed state using single QD.
\section{Acknowledgements}
This work was supported by DST SERB Fast track young scientist scheme SR/FTP/PS-122/2011.

\end{document}